\documentclass{acm_proc_article-sp}

\usepackage{amssymb}
\usepackage{amstext}
\usepackage{amsmath}
\usepackage{algorithm2e}
\usepackage{url}
\usepackage{subfigure}
\usepackage{graphicx}

\begin{document}

\title{Collaborative Filtering without Explicit Feedbacks for Digital Recorders}

\numberofauthors{4} 
%
\author{
%
%
Alessandro Basso, Marco Milanesio, Andr\'e Panisson, Giancarlo Ruffo\\
       \affaddr{Dipartimento di Informatica}\\
       \affaddr{Universit\`a degli Studi di Torino}\\
       \affaddr{Torino, Italy}\\
       \email{\{basso,milane,panisson,ruffo\}@di.unito.it}
}


\maketitle
\begin{abstract}
Recommendation is usually reduced to a prediction problem over the function $r(u_a, e_i)$ that returns the expected rating of element $e_i$ for user $u_a$. In the IPTV domain, we deal with an environment where the definitions of all the parameters involved in this function (i.e., user profiles, feedback ratings and elements) are controversial. To our knowledge, this paper represents the first attempt to run collaborative filtering algorithms without inner assumptions: we start our analysis from an unstructured set of recordings, before performing a data pre-processing phase in order to extract useful information. Hence, we experiment with a real Digital Video Recorder system where EPG have not been provided to the user for selecting event timings and where explicit feedbacks were not collected.
\end{abstract}




\section{Introduction and Context}\label{intro}

In the wide context of IPTV services, \textit{Digital Video Recorders (DVR)} (a.k.a. \textit{Personal Video Recorder (PVR)}), are hardware or software devices that record digital video to a memory medium, for a further access (e.g., stand-alone set-top-boxes (STB), portable media players, and PC based DVRs). One of the most important exploitations of the DVRs is the \textit{media sharing}: users may want to access to all their personal media resources (e.g., pictures, podcasts, TV programs) from any of their own devices (e.g., laptops, smartphones and so on). For example, Orb\footnote{http://www.orb.com} is a freeware streaming software that enables users to access remotely their media via Internet. In this paper, while focusing on the recording of radio and TV broadcasts, we keep in mind the general media sharing domain.

Recording and personalization is changing the way providers insert advertisements during content delivery. The emergence of new behaviors and content consuming trends is forcing advertisers to look for new ways (e.g., a pull model) for spreading their commercials, trying to avoid the fast-forward of pre-recorded videos to skip commercials. In this domain, thus, \textit{recommender systems} \cite{adomavicius05toward} are becoming worthwhile as retention tools for customers. When a user is satisfied with suggestions, she dedicates more attention to the proposed links and references. Even if some recommendation engine has been tailored for DVRs (e.g., Neptuny's Content Wise, ReignSoft's Impress), a comprehensive analysis of the peculiarities of this domain is still missing. 

First of all, usage data collection is subject to serious privacy concerns. Users are not willing to loose control of data they produce while taking advantage of recommendations and personalized information. Moreover, when a shared device is used (e.g., the television) also family control issues arise, and it can be difficult to provide personalized information. Even for such reasons, costumers protect themselves behind fictitious on line identities, and this is an important challenge for many recommendation algorithms, that need \textit{user profiles} to increase their accuracy.\\
Another important problem in the IPTV domain (\cite{cremonesi09analysis,cremonesi09recommender}) is that, differently from other domains where recommenders can learn from explicit user ratings over content, in DVR systems we need to operate according to \textit{implicit feedbacks}, such as recordings of a given event, downloading of pre-recorded videos, or - when possible - monitoring if the video has been effectively watched by the user. \\
A third relevant problem, underestimated by previous works on this subject, is the difficulty of discriminating items. DVRs usually provide Electronic Program Guides (EPGs) to help users. Unfortunately, in a wider context, EPGs are not reliable to identify recommendable elements: TV channels may not respect schedules, media and podcasts available on the Web do not use a common schedule's format, users may be interested only in single parts of an event, and thus they will set up their own timings over a given channel. In opposition with VoD, content description is difficult in the TV domain. Broadcasts are announced, but often lack of structured meta-information: thus, collaborative filtering is usually preferred to content-based recommendation algorithms, even if its execution is not straightforward.

Recommendation is usually reduced to a prediction problem over the function $r(u_a, e_i)$ that returns the expected rating of element $e_i$ for user $u_a$. As observed above, we are dealing with an environment where the definitions of all the parameters involved in this function (i.e., user profiles, feedback ratings and elements) are controversial. To our knowledge, this is the first attempt to run collaborative filtering algorithms without inner assumptions.

In this paper, we start our analysis from an unstructured set of recordings, before performing a data pre-processing phase in order to extract useful information. Hence, in Section \ref{context} we present a real system without EPGs where explicit feedbacks are not collected. Then, we describe the procedure for extracting meaningful information from the large and unstructured amount of provided data  (Section \ref{extraction}) and the analyzed recommendation algorithms (Section \ref{rec}). Finally, the evaluation of the chosen algorithms is presented in Section \ref{results}, before drawing conclusions.

\section{The Faucet PVR Environment}\label{context}

Our analysis is based on real data generated by the \textit{Faucet PVR} system, integrated in a web-based podcasting service named \textit{VCast}\footnote{\url{http://www.vcast.it/}}. Faucet allows users to record their favorite (Italian) TV and Radio programs, and to further download them  into their devices (e.g., iPod, PC, notebook) \cite{carmagnola2009vcast}. The user can set up her own programming and see or download her recordings by the use of a simple web interface. Bringing the ability to record and group into a single feed public and private channels (such as radio and TV recorded programs), Faucet PVR offers a single framework for creating and aggregating personal podcast compilations.

The Faucet PVR produces a very rich and dynamic dataset\footnote{The Vcast dataset is publicly available at: \url{http://secnet.di.unito.it/vcast}}, populated by real users expressing their preferences through the recorded programs. Such a context, however, is characterized by a number of constraints which we had to deal with, in order to be able to perform the analysis on the recommendation algorithms. In particular, the intrinsic dynamism and variability of the recordings, as well as the lack of any permanent event, require that a series of pre-processing steps have to be undertaken prior to be able to apply any recommender.

A noticeable property characterizing the context in which we operate is the lack of a well defined programming for recorded contents. Despite several EPG sources do exist, we can not consider them reliable enough to be used for extracting the input information of the recommender engine. Therefore, we decided to opt for a different approach. Exploiting a bottom up approach, we rely on users' knowledge to define the most relevant properties of the events transmitted on the major TVs and radios.\\
More precisely, the task of defining the parameters related to every recorded transmission is therefore demanded to single users. Since it is their primary interest to make sure that the information inserted in the Faucet PVR are as much precise as possible, we can consider such data reliable enough to be used in the recommendation process. Furthermore, a number of inferences can be deduced from the user activity, and considering also the good popularity of the system, also numerical processing is statistically reliable. 

The Faucet PVR involves three different steps to be taken by an user when she is interested in recording an event: the \textit{parameters setting}, the \textit{execution} and the \textit{downloading} of the recorded item. All three steps are performed in different moments, in the aforementioned order. In the parameters setting step, the user chooses a channel, periodicity, name, starting and ending times. This step must be done before the beginning of the program. The execution phase starts at the starting time and finishes at ending time. The downloading step is available only after the recording finishes.

As mentioned above, an event can be periodic: if the user wishes, the system records the desired event in regular intervals. These intervals can be of a week or a day, and in the case of a daily event, the user can choose to skip weekends. Events classified as non-periodic have absolute starting and ending times. However, in the case of periodic events, starting and ending do not represent absolute times, but rather a weekday and daytime (in the case of weekly events) or just a daytime (in the case of daily events).
Also, in the case of periodic events, there is one parameter setting step, but the execution step can occur an undefined number of times. After each execution step, a download referring to it is made available. The system limits the number of accumulated recordings to 3 in order to save resources (only the last 3 executions are available to download).

The fact that the dataset includes information about periodicity implies some issues in properly determining the events broadcasted on TV and radio from the amount of recordings made by users. On the other side, it decreases the complexity of calculating recommendations, resulting in an overall improvement in their novelty. In fact, without taking into account the periodicity, as in \cite{koren08collaborative}, the recommender has to explicitly ignore periodic elements recently seen by the user, in order to provide a more valuable and accurate recommendation. In our domain, as the periodicity is an intrinsic feature of the recommendable items, we do not have such a constraint, being these elements automatically excluded.

The goal of a recommendation system in the PVR context is to suggest a personalized set of transmissions to the users. However, data coming from the Faucet PVR are not immediately usable to identify events such as the transmissions, but assume the form of unstructured information, which have to be properly processed. In particular, let $T$ be the set of transmissions during a day and $t_i$ be a specific transmission broadcasted on channel $c_{t_i}$, starting at time $b_{t_i}$ and ending at time $e_{t_i}$. Then, $t_i$ can be directly used in the recommendation engine, as well as $\forall t \in T$.
On the contrary, data collected by the Faucet system (i.e., users' recordings) differ from $t_i$ in the sense that they define a set $R$ of several events $r_i$ with a temporal validity, each referring to a specific event. However, recordings with different timings may refer to the same broadcast: given the pair $r_i, r_j \in R$, they may refer to the same transmission even if $r_i \neq r_j$. Clearly, this property does not hold with discrete and well defined events such as the transmissions. 

As well as we can not exploit any EPG to identify the recorded contents, we can not even rely on any information about the specific content of each recording. Indeed, the Faucet PVR does not provide any reference to the type of transmission recorded (e.g., sport program, news, or comedy-movie), nor we can rely on information inserted by users in \textit{title} field, as the insertion of titles and annotations is completely free, and this results in very diversified and, possibly, incorrect descriptions.

A further implication due to the lack of information about the content of the recordings is the impossibility of applying content based recommenders, which focus on the description of item and user profiles in order to recommend items \cite{citeulike:1668956}. On the contrary, our approach relies only on the behavior and characteristics of large interconnected networks, by exploiting relations between their users. Each specific user follows her personal \textit{behavioral pattern} in the usage of the Faucet PVR, and we can investigate these patterns to compute a similarity among users.

As a final observation, broadcasting is characterized by the \textit{expiration} of some events: we can suggest the user to record only future broadcasts, and even if some shows are serialized, the recording of the single episode should be programmed in advance. This phenomenon is (partially) due to copyright management, since the content provider are not willing to authorize service providers to store previously recorded event for further distribution. Nevertheless, recording of a broadcast is still allowed, because it is seen as a single user activity. As a consequence, we have to deal (also) with volatile content, and this differs very much with the VoD domain, that has been exhaustively explored in the context of recommendation systems.

\section{Data Extraction}\label{extraction}

Due to the specific domain, we are required to perform a pre-process of the data obtained from the Faucet PVR prior to be able to use such information as input for a recommendation algorithm. This is needed because the Faucet system generates a set of recordings in the continuous domain of \textit{timings}, while a recommender system requires to operate in the discrete domain of \textit{events}.

As a consequence, the first goal that we have to accomplish is the identification of the broadcasted transmissions from the amount of unstructured data resulting from the recording process. This is a multi-step procedure, whose aim is to identify a set of \textit{discrete elements} as the representatives of the broadcasted \textit{events}. Basically, a discrete element is obtained as the result of the aggregation of several different recordings. A preliminary investigation on the extraction of events from recordings is given in \cite{basso09events}.

Let $U = \{u_1, u_2, ... , u_k\}$ be the set of distinct users in the Faucet platform. Each user in set $U$ has recorded some programs in the past and scheduled some for the future. To schedule a program, a user must choose a channel $c \in C$, representing a list of predefined channels, and a periodicity $p \in \{no-repeat, weekly, daily, mon-fri, mon-sat\}$, representing all the possible periodicities allowed in the PVR system. Besides, the user is required to annotate his/her recording with a (possibly) meaningful title.

Let $R = \{r_1, r_2, ... , r_m\}$ be the set of the broadcasted recorded programs. Each element in $R$ (a recording) is a tuple $r_i = <u_i, c_i, p_i, t_i, b_i, f_i>$ set by a user $u_i \in U$ who recorded on the channel $c_i \in C$ with periodicity $p_i \in P$ a program titled $t_i$ with start time $b_i$ and end time $f_i$. Thus, we can assume that there exists a function mapping every user to her recordings.

The set $R$ is first processed by means of clustering; then, aggregation and collapsing are carried out in sequence on the output of the clustering. The three phases are described in the following.

\paragraph*{Clustering}\label{sub:clustering}
Due to the lack of information about the content of each recording, they are clustered wrt the channel, the periodicity and the difference between timings. Specifically, $\forall{r_i,r_j} \in R | c_{r_i} = c_{r_j} \wedge p_{r_i} = p_{r_j}$ we have that
\[ r_i \biguplus r_j \mbox{ iff } |b_{r_i} - b_{r_j}| < \delta_b \wedge |f_{r_i} - f_{r_j}| < \delta_f, \]
where $\biguplus$ is the clustering function and $\delta_b, \delta_f$ determine the maximum clustering distance for the start and end times, respectively. The identified clusters contain recordings equal in the channel and periodicity, and similar on the timing. The recording that minimizes the intra-cluster timing distances is elected as the centroid of the cluster. At the end of the clustering, each cluster identifies an event.

\paragraph*{Aggregation}\label{sub:aggregation}
As the Faucet platform produces new recordings with a hourly frequency, we perform the clustering once a hour obtaining a set of newly generated events. A further step is then required to possibly aggregate similar events, i.e., the new one with those previously created. Such an operation is performed as follows: (1) we compare each element generated with the clustering with the existing events wrt channel, periodicity and timings; (2) if the timings are similar, we correct the properties of existing events with the values of the newly created ones. The list of the users associated to the event is updated accordingly.

\paragraph*{Collapsing}\label{sub:collapsing}
Similar discrete elements, i.e. with the same channel and periodicity but timings within a fixed range, are merged into a single event. All features of the new events are computed by means of the values of the collapsed discrete elements. This operation is required basically because events can be created in subsequent moments, by aggregating recordings referring to the same broadcasted transmissions. Due to the high variability of the timings, especially when a new transmission appears, such events slowly and independently converge to more stable timeframes, determining the need of collapsing them into single events.

As a result of the whole process, we obtain a number of events, each being a tuple defined as follows:
\[e_j = <\{u_{e_j}\}, c_j, tl_j, b_j, f_j, p_j>\]
where:
\begin{itemize}
  \setlength{\itemsep}{0pt}
  \setlength{\parsep}{0pt}
  \setlength{\topsep}{0pt}
  \setlength{\partopsep}{0pt}
  \setlength{\parskip}{0pt}
  \setlength{\labelwidth}{0em}
  \setlength{\labelsep}{.25em}
  \item $\{u_{e_j}\}$ is the list of users who set a recording referring to that event;
  \item $c_j$ is the channel;
  \item $tl_j$ is the title chosen among those given by users;
  \item $b_j$ and $f_j$ are the the starting and ending times respectively;
  \item $p_j \in \{no-repeat, weekly, daily, mon-fri, mon-sat\}$ is the periodicity.
\end{itemize}

\begin{figure}[!htb]
 \centering
 \includegraphics[scale=0.53, keepaspectratio]{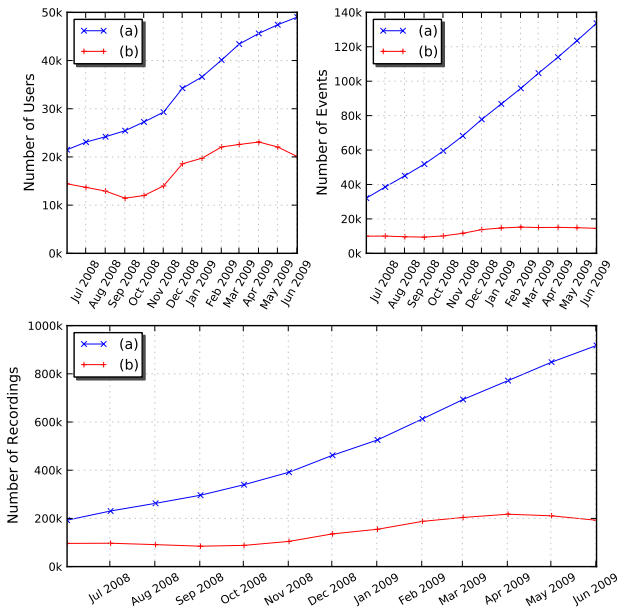}
 \caption{Number of users, events and recordings (a) Total and (b) Active in the considered period}\label{fig:user_item_recording_number}
\end{figure}

In Figure \ref{fig:user_item_recording_number}, we can observe the behavior of the system in a one year timeframe, i.e., from June 2008 to June 2009, wrt the number of users, events and recordings. As the number of active recordings and events (b) tends to increase over time, the number of users follows a different, less constant, trend. Specifically, we can notice a considerable increase in the number of users in the system between November 2008 and March 2009. Such a happening implies a consequent raise in the number of recordings, due to the augmented activity in the system. Analogously, the number of events generated by the aggregations of the recordings grows up, although less noticeably if compared to the recordings.

\section{Recommendation}\label{rec}

In our context, we can identify two different approaches to recommendation, depending on the specific target which is considered. As a first attempt, we define a set of events which can be of interest to the majority of the users. In such a case, we are trying to identify the \textit{most frequent} events in the systems, i.e., those programs which have been recorded by the largest subset of users. This is done by means of a recommendation algorithm which we name \textit{MostPopular}.\\
A second approach focuses on identifying those events which can be of any interest for a single user of the system. In this case, which we can refer to as user-oriented, the aim of the recommendation algorithm is to suggest items specifically tailored to the user's preferences.

\subsection{Algorithms Overview}\label{rec:algorithms}
Two well-known recommendation techniques are considered in this work: (1) the memory based \textit{collaborative filtering} approach named \textit{k}-Nearest Neighbors (kNN) \cite{sarwar01item-based}; (2) the model based approach based on the \textit{SVD transform} \cite{sarwar00applicationof}.\\
Exploiting the basic idea of the \textit{nearest neighbors} approach, we apply both variants of the kNN algorithm: the user-based one \cite{herlocker99algorithmic}, by identifying users interested in similar contents; and the item-based approach \cite{deshpande04item-based}, by focusing on items shared by two or more users. In addition, we also analyze the performance of a variant of the SVD technique based on implicit ratings, presented in \cite{koren08collaborative}.

\paragraph*{User-based kNN}
In the \textit{user-based} kNN algorithm, the weight of an element $e_i$ for an user $u_k$ can be defined as:
\begin{equation}
\label{eq:knnweight}
\mbox{w}(u_k,e_i) = 
\sum\limits_{u_a \in N(u_k)}  \mbox{r}(u_a,e_i)\cdot\mbox{c}(u_k, u_a), 
\end{equation}
\[
\text{where} \quad \mbox{r}(u_a,e_i) = 
\left\{ 
\begin{array}{l l}
  1 & \text{if} \quad e_i \in E_{u_a}\\
  0 & \text{if} \quad e_i \notin E_{u_a}\\
\end{array} \right.
\]
$E_{u_a}$ is the set of elements recorded by user $u_a$, whilst $N(u_k)$ is the neighborhood of user $u_k$, limited by considering only the top-\textit{N} neighbors ordered by user similarity. The different similarity functions are discussed in section \ref{rec:knn:neighborhood}. In case the number of neighbors is limited by the chosen similarity function to a number lower than $k$, we also consider the 2nd-level neighbors, i.e., for each user $u_a$ belonging to $N(u_k)$ we compute $N(u_a)$. The overall set of 1st-level and 2nd-level users is then used to define the users similar to $u_k$, as previously described.

The coefficient $\mbox{c}(u_k, u_a)$ represents the neighbor's information weight for user $u_k$. In most of the kNN-based algorithms \cite{herlocker99algorithmic}, the coefficient used is the similarity between $u_k$ and $u_a$. In other cases \cite{bell07scalable} the coefficients are calculated using derived interpolation weights. 
It is worth noting that, in case of considering 2nd-level neighbors, the coefficient $\mbox{c}(u_k, u_a)$ in eq. \eqref{eq:knnweight} has to be computed taking into account the similarity between the considered neighbor and further ones. For example, considering user $u_k$, her neighbor $u_a$ and her 2nd-level neighbor $u_b$, we have:
\[
\mbox{c}(u_k, u_b) =  \mbox{c}(u_k, u_a) * \mbox{c}(u_a, u_b),
\]
that is a combination of the similarities computed between the neighbors pairs for the considered user.

\paragraph*{MostPopular}
The \textit{MostPopular} algorithm can be also defined by means of eq. \eqref{eq:knnweight}, assuming the number of neighbors unbounded, which implies $N(u_k) = U, \ \forall u_k \in U$; and $\mbox{c}(u_a, u_b) = 1, \ \forall u_a,u_b \in U$.\\
The weight of an element $e_i$ to an user $u_k$ is therefore defined as:
\begin{equation}
\label{eq:mostpopular}
\mbox{w}(u_k,e_i) = 
\sum\limits_{u_a \in U} \mbox{r}(u_a,e_i)
\end{equation}
After calculating the weight of all elements, they are sorted in descendant order. In the \textit{MostPopular} algorithm, as the set of neighbors is independent of the user, all users receive the same recommended elements, i.e., the most popular elements.

\paragraph*{Item-based kNN}
In the \textit{item-based} \textit{k}NN algorithm, the weight of an element $e_i$ for an user $u_k$ is defined as:
\begin{equation}
\label{eq:itembasedknnweight}
\mbox{w}(u_k,e_i) = 
\sum\limits_{e_j \in N(e_i)}  \mbox{r}(u_k,e_j)\cdot\mbox{c}(e_i, e_j), 
\end{equation}
$N(e_i)$ is the set of $n$ items most similar to $e_i$ and recorder by $u_k$, and $\mbox{c}(e_i, e_j)$ is the neighbor's information weight wrt item $e_i$.

Differently from the user-based case, using $k=\infty$ in the item-based approach does not lead to the \textit{Most Popular} set of elements. In fact, the algorithm simply takes all items $e_j \in E_{u_k}$ as neighbors of $e_i$, making $N(e_i)$ user-dependent.

\paragraph*{SVD}
The Singular Value Decomposition technique analyzed in this work makes use of implicit feedbacks and implements the method proposed in \cite{koren08collaborative}. Specifically, given the observations of the behavior of user $u$ wrt item $i$, $r_{ui}$, we can define the user's preference as:
\[
  p_{ui} = 
  \left\{ 
  \begin{array}{l l}
    1 & \text{if} \quad r_{ui} > 0\\
    0 & \text{if} \quad r_{ui} = 0\\
  \end{array} \right.
\]
where $r_{ui}$ is set to 1 when $u$ records item $i$, 0 otherwise.

After associating each user $u$ with a user-factors vector $x_u \in \mathbb{R}^f$ and each item $i$ with an item-factors vector $y_i \in \mathbb{R}^f$, we can predict the unobserved value by user $u$ for item $i$ through the inner product: $x^{T}_{u}y_i$.
Factors are computed by minimizing the following function \cite{koren08collaborative}:
\[
  \min_{x_{\text{*}}y_{\text{*}}} \sum_{u,i} (p_{ui} - x^{T}_{u}y_i)^2 + \lambda \left( \sum_{u}\lVert x_u \rVert^2 + \sum_{i}\lVert y_i \rVert^2 \right) 
\]

\subsection{Computing Neighborhood}\label{rec:knn:neighborhood}
In order to provide recommendation on the discrete elements, we have to define a similarity function for grouping similar users/items from which choosing the appropriate elements to recommend. The definition of the similarity is based only on implicit ratings resulting from observing the behavior of users: if she records something, then we assume that she is interested in it; otherwise, we can not infer anything about the interest of the user for that element. We are therefore considering binary feedbacks.

\paragraph*{User-to-user}\label{rec:knn:neighborhood:user2user}
Let $u$ and $v$ be two users and $E_u, E_v$ the sets of recorded elements associated to them; we can choose the similarity metric, $S(u, v)$, considering several well known measures, such as: the \textit{Jaccard}'s coefficient, the \textit{Dice}'s coefficient, the \textit{Cosine} similarity and the \textit{Matching} similarity \cite{markines09evaluating}.\\
Then, $\forall u $, we can then compute the subset $N_u \subseteq U$ of \textit{neighbors} of user $u$. A user $v$ such that $E_v \cap E_u \neq \emptyset$ is thus defined as a neighbor of $u$. Starting from the neighborhood of $u$, similarity with $u$ is computed for each pair $<u, v>$ such that $v \in N_u$.\\
Finally, if $S(u, v) > 0$, we consider $u$ similar to $v$, i.e., there is an arc connecting them in the similarity network. The value $S(u, v)$ is used to weight such a relation, therefore determining a similarity order among the neighborhood of $u$.

\paragraph*{Item-to-item}\label{rec:knn:neighborhood:item2item}
The similarity among items, $S(e, f)$, is based on the same measures already mentioned before, yet redefined considering two items $e,f$ and their sets of users $U_e, U_f$ who recorded them.\\
$\forall e \in E $ we can compute the subset $N_{e} \subseteq E$ of \textit{neighbors} of item $e$. An item $f$ such that $U_{e} \cap U_{f} \neq \emptyset$ is thus defined as a neighbor of $e$. Starting from the neighborhood of $e$, similarity with $e$ is computed for each pair $<e, f>$ such that $f \in N_{e}$.\\
We can then decide whether a couple of items is similar or not. Items $e$ is considered similar to $f$, i.e., there is an arc connecting them in the similarity network, if $S(e, f) > 0$. A similarity order among the neighbors of $e$ is thus determined.

\section{Experimental Results}\label{results}

Our evaluation is based on trying to measure how accurate is each recommendation algorithm in predicting the elements that users would program. This is achieved by computing precision and recall on the predicted items. The more accurate is this prediction, the more valuable elements are recommended. It is important to underline that we do not consider any feedback related to the user's interest in the recommended items, but we only focus on the prediction ability of the algorithms analyzed.

To start evaluating a recommendation algorithm, we fix an arbitrary time $t$ after the data collection started and before the data collection stopped.  The value of $t$ should be carefully chosen not to be too close to the data collection start, since we do not have sufficient data to make good predictions. Also, the time $t$ should not be close to the end of data collection, because we need a good amount of data to make the verification if the algorithm was able to predict it. As the data collection started January 23rd 2008 and ended November 19th 2009, we choose values of $t$ varying from June 1st 2008 to June 1st 2009.

\subsection{Metrics}\label{results:metrics}
Given the set $E$ of events in our framework, we define the following subsets:
\begin{itemize}
  \setlength{\itemsep}{0pt}
  \setlength{\parsep}{0pt}
  \setlength{\topsep}{0pt}
  \setlength{\partopsep}{0pt}
  \setlength{\parskip}{0pt}
  \setlength{\labelwidth}{0em}
  \setlength{\labelsep}{.25em}
  \item $A(t) \subset E$, active events at time $t$ ($b_j > t$);
  \item $R(u,t) \subset E$, events recorded by user $u$ before time $t$;
  \item $V(u,t) \subset A(t)$, events recorded by user $u$ after time $t$;
  \item $Rec(u,t) \subset A(t)$, events recommended to user $u$ at time $t$.
\end{itemize}
It is important to notice that $A(t)$ is also the set of all elements suitable for recommendation at time $t$. The aim of our recommendation algorithms is to predict which events are in $V(u,t)$. For that, for each user, the algorithms associate a weight $w(u,s)$ to each element $s \in A(t)$ which represents, from the recommender's point of view, how much reliable is the fact that $s \in V(u,t)$. Furthermore, a recommendation algorithm use only the information in
\[\bigcup_{u \in U} R(u,t), \quad\text{with}\quad R(u,t) \cap V(u,t) = \emptyset.\]
To recommend items to users, we use only the top $n$ recommended elements $Rec(n,u,t) \subset Rec(u,t)$, i.e., the top $n$ elements in $Rec(u,t)$, ordered by weight.
The \textit{precision} and \textit{recall} at time $t$ are computed as the average of all users' precision and recall values computed using the top $n$ recommended elements \cite{sarwar00applicationof}. Finally, we compute the system precision and recall at different times, and calculate the system overall precision and recall as the average of it.

As in \cite{koren08collaborative}, also in our context precision measures are not very meaningful, because we do not have feedbacks regarding the user's interest in those items which have not been considered (i.e., not programmed, nor downloaded).
On the contrary, recall-oriented measures are more suitable. Infact, we can assume that $e_i$ is of any interest for user $u$ only if $e_i \in V(u,t)$, otherwise no assumption on user's interests can be made. Anyway, for sake of completeness, we also report the analysis of precision values.

\subsection{Evaluation}\label{results:evaluation}
\begin{figure*}[!htb]
 \centering
 \subfigure[Upper and lower bounds for recall]{
  \label{fig:upper_lower_bounds}
  \includegraphics[scale=0.35, keepaspectratio]{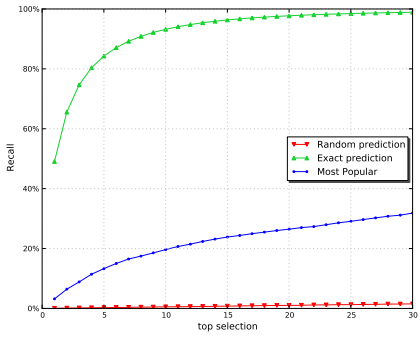}
 }
 \subfigure[Comparison between similarity functions in user-based kNN]{
  \label{fig:userbased_similarity_comparison}
  \includegraphics[scale=0.35, keepaspectratio]{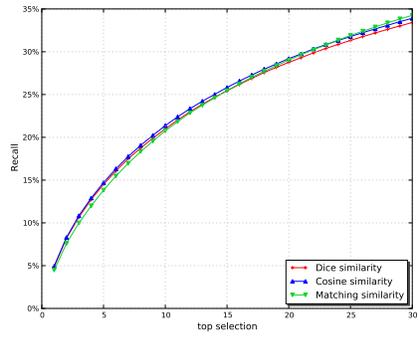}
 }
 \subfigure[Comparison between similarity functions in item-based kNN]{
  \label{fig:itembased_similarity_comparison}
  \includegraphics[scale=0.35, keepaspectratio]{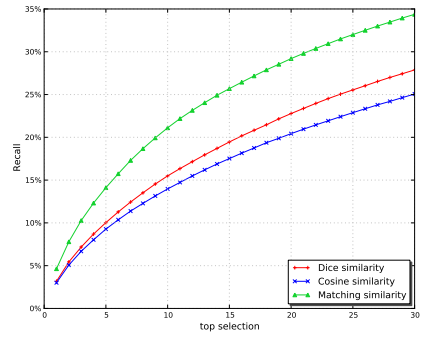}
 }
 \caption{Comparison between recommenders and similarity functions.}
\end{figure*}

As a first step in the evaluation, we attempt to define the specific upper and lower bounds which characterize the recommendations in the PVR domain. In particular, we compare the \textit{MostPopular recommender}, which identifies the most frequent elements among all users (Section \ref{rec:algorithms}), with the following two algorithms: (1) a \textit{random recommender}, which simply chooses $n$ random elements among those of $A(t)$, defining the lower bound to our experiment; (2) an \textit{exact predictor recommender}, which has knowledge about the elements in $V(u,t)$, thus yielding to the best possible results and defining the upper bound.\\
The results are depicted in Figure \ref{fig:upper_lower_bounds}, which clearly shows that, as expected, even the \textit{MostPopular} algorithm can easily outperform a random predictor. However, it is still far from being able to make a complete prediction of all the elements, especially when the considered top \textit{n} are just few items.

The second step in our evaluation is to study how different similarity functions affect the results of user-based $k$NN recommendation algorithms. 
We can observe from Figure \ref{fig:userbased_similarity_comparison} that, in case of the user-based algorithm, all chosen similarities show nearly the same performances.\\
On the contrary, the Matching similarity considerably outperforms the other measures when it comes to the item-based algorithm, as displayed in Figure \ref{fig:itembased_similarity_comparison}. Again, both Dice and Jaccard show a very similar behavior, being superior to the Cosine metric already when more than 5 elements are recommended. In both Figures \ref{fig:userbased_similarity_comparison} and \ref{fig:itembased_similarity_comparison}, the Jaccard similarity is not shown being almost identical to the Dice.

\begin{figure*}[!htb]
 \centering
 \subfigure[Comparison of one-level and two-level neighborhoods for user-based $k$NN]{
  \label{fig:userbased_neighborhood_level_comparison}
  \includegraphics[scale=0.35, keepaspectratio]{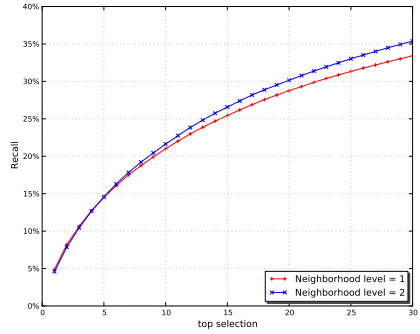}
 }
 \subfigure[Recall for user-based kNN]{
  \label{fig:recall_userbased}
  \includegraphics[scale=0.35, keepaspectratio]{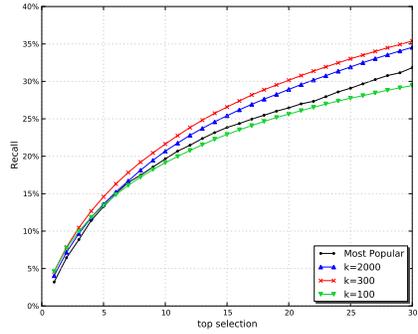}
 }
 \subfigure[Precision vs Recall for user-based kNN]{
  \label{fig:precision_recall_userbased}
  \includegraphics[scale=0.35, keepaspectratio]{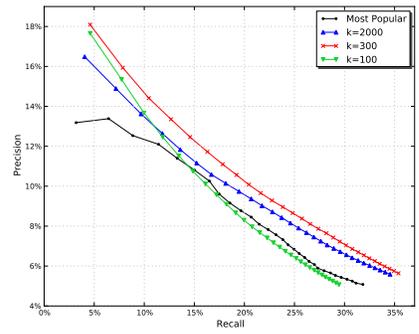}
 }
 \caption{Neighborhoods comparison, precision and recall for user-based $k$NN.}
\end{figure*}

Another step in our evaluation is to find the consequences of adding second-level neighbors in the neighborhood of user-based $k$NN recommendation algorithms. In Figure \ref{fig:userbased_neighborhood_level_comparison}, we can observe that increasing the number of first level neighbors (when it is lower than $k$) by adding the second level ones implies a better performance of the algorithms. In this example, we used Dice similarity and $k=300$, however the results are similar when applying second-level neighbors to other similarities.

In the next tests, we try to find an optimal value for $k$ in the user-based $k$NN algorithm. Figure \ref{fig:recall_userbased} shows the results of $k$NN user-based with $k \in \{100, 300, 500, 700, 2000\}$, and the \textit{MostPopular} recommender. We used Dice similarity, but the results are similar with other similarity functions. In addition, in Figure \ref{fig:recall_userbased}, as well as in Figure \ref{fig:precision_recall_userbased}, we omit the values of $k=\{500,700\}$ since the results are very similar to the case of $k=300$.\\
We can observe that a value $k=100$ is not sufficient to outperform the \textit{MostPopular} algorithm, due to the lower value of the recall. On the other side, a very high number of neighbors allows to perform better than the \textit{MostPopular}. However, we could notice that, already with $k=2000$, the algorithm starts to converge to the \textit{MostPopular}, characterized by an unbounded number of neighbors by definition. Therefore, we can consider the range $[100, 2000]$ as suitable to identify the optimal value for $k$.\\
For this purpose, we test the values $k=\{300, 500, 700\}$, obtaining very similar performance. Considering the top 10 recommended elements, we can achieve better results for $k=300$, whilst $k=500$ is more suitable when taking the top 11 to 30 elements. As in most cases 10 elements are sufficient for a recommendation, $k=300$ is a good compromise between the ability of providing valuable recommendations and the resource consumption in calculating the neighborhood.

To better observe the trend of both recall and precision, Figure \ref{fig:precision_recall_userbased} shows the two values combined. Again, $k=300$ performs better if we take only the top 10 recommended elements, as it also yields to good results in terms of precision. Considering more than 10 recommendations, it would seem appropriate to increase the number of neighbors to 500, as the results for precision and recall are slightly better. However, the overall behavior of the algorithm is almost identical with $k$ in the range $\{300,700\}$. Nevertheless, the above mentioned considerations regarding the superior performance of the kNN algorithm with $k=300$ in terms of computation requirements still apply when we take into account the precision metric.

\begin{figure*}[!htb]
 \centering
 \subfigure[Recall for kNN ($k=300$) wrt MostPopular]{
  \label{fig:userbased_itembased_comparison}
  \includegraphics[scale=0.35, keepaspectratio]{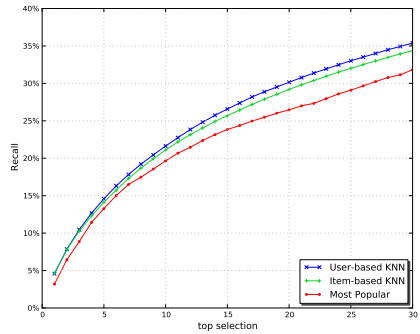}
 }
 \subfigure[Recall for SVD wrt kNN ($k=300$) User-based and MostPopular]{
  \label{fig:recall_svd}
  \includegraphics[scale=0.35, keepaspectratio]{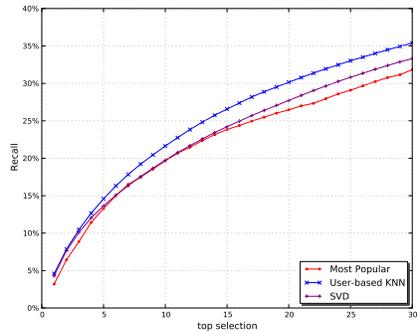}
 }
 \subfigure[Precision vs Recall for kNN ($k=300$), SVD and MostPopular]{
  \label{fig:userbased_itembased_precisionrecall_comparison}
  \includegraphics[scale=0.35, keepaspectratio]{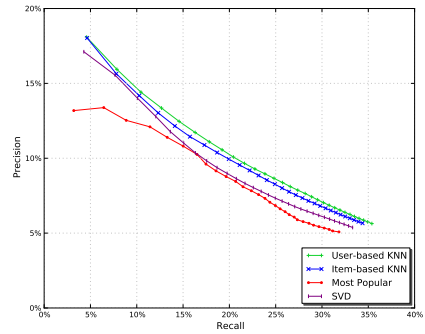}
 }
 \caption{Precision and recall for the analyzed algorithms.}
\end{figure*}

An interesting comparison among the three kNN algorithms analyzed, i.e., user-based, item-based and \textit{MostPopular}, is depicted in Figure \ref{fig:userbased_itembased_comparison}. We can observe that the latter is clearly outperformed by the other two algorithms in terms of recall, especially when more than 7 recommended items are considered. Between the item-based and the user-based version of the kNN, the latter performs slightly better, although the gap is mostly noticeable when more than 15 items are recommended. In general, item-based algorithms tend to perform better because usually the number of items is considerably lower than the users \cite{sarwar01item-based}. Such a property does not hold in our domain, hence making the user-based version superior in terms of recall, as we initially expected.

A final experiment is attempted in order to measure the behavior of the SVD approach wrt the performance of the kNN method. The implementation of the SVD algorithms described in Section \ref{rec:algorithms} is tested with different parameters, with the purpose of identifying the more suitable ones in our context. In particular, we try different sizes for user-factors and item-factors vectors, values for the $\lambda$ parameter and number of training steps.\\
Results are depicted in Figure \ref{fig:recall_svd}. The best prediction is obtained with 100 features, $\lambda = 500$, $\alpha = 40$ and 15 training steps. However, the behavior of the latent factor model based on SVD in the analyzed context is worse if compared to a neighborhood model such as kNN. As the reader can notice, the kNN user-based is able to substantially outperform the SVD technique, whose results in terms of recall are slightly better than those of the MostPopular algorithm only when a considerable number of items are recommended.

Similarly, results related to the precision of the recommendations (Figure \ref{fig:userbased_itembased_precisionrecall_comparison}) show an analogous behavior of the kNN algorithms wrt SVD, with the \textit{Most Popular} being considerably less precise than others. Also, the user-based algorithm shows to be more precise than the item-based in determining the recommendable items, for the same reason previously mentioned considering recall.

It could appear surprising that the prediction performance of the SVD recommender is worse than other techniques, as this algorithm normally performs better in several other contexts \cite{sarwar00applicationof, koren10collaborative-cacm, koren08collaborative}. We believe that the motivations for such an unusual behavior reside in the dataset characteristics. In particular, a reason might be identified in the so called \textit{cold start problem}, whose effects involve users, items and communities \cite{schafer99}.\\
In our context, the cold start problem is particularly noticeable with items and is due to the lack of relevant feedbacks when a new event first appears in the system. Such an issue is made worse by the fact that items to recommend are generally new ones, i.e. those events having a starting time in the future. This property holds for no-repeat events as well as for repetive ones (the starting time is updated according to their periodicity). So, events whose starting time has passed are no longer elegible for recommendation.

The fact that recommendations are affected by the cold start problem is one key factor that may influence SVD performance, as this algorithm needs support of user's preferences to perform well. On the contrary, a neighborhood-based approach such as kNN appears to better deal with newly introduced items, as also reported in \cite{cremonesi09analysis}.

\section{Conclusion}\label{conc}

We experimented with a real digital recording service, and, accordingly to the above mentioned restrictions, we decided to run our analysis under the strongest assumptions: no EPGs are available, users can set up timings as well as channels, explicit feedbacks are not collected, and so on.
In addition, the intrinsically dynamic nature of the analyzed PVR domain, which determines a continuous process of creation and deletion of events and a consequent amplification of the cold start problem, makes such a context sensibly different in terms of recommendation if compared to those where items have no time validity (i.e., netflix, movielens, etc.).

Despite these constraints, our results showed that neighborhood based strategies, such as kNN, can return in good prediction accuracy and, if correctly tuned, they can outperform SVD-based techniques as well as \textit{most popular} strategies, that dangerously leverage the phenomenon of many users concentrated on very few relevant events.\\
Finally, there is evidence that digital recorders differ from other interest based services, because factors other than personal tastes might influence the user's behavior and the success of a recommendation engine. In fact, the direct social influence of friends and the volatile nature of events are supposed to be relevant factors in causing a user to schedule a recording.\\
In our opinion, a possibile future research direction could be indeed the identification and study of those social factors which affect user's behavior in systems characterized by high dynamism and short lifetime of items.

%
\bibliographystyle{abbrv}
\bibliography{rdpvr,recommender}  
%
%

\balancecolumns
\end{document}